\documentclass[prl,aps,floats,amssymb,twocolumn,preprintnumbers,footnoteinbib]{revtex4}

\usepackage{amssymb}
\usepackage{epsfig}

\newcommand{\gev}{\, {\rm GeV}}

\newcommand{\fm}{\, {\rm fm}}

\newcommand{\Rlsd}{{}^{\ell}R_d^s}

\newcommand{\be}{\begin{equation}}
\newcommand{\ee}{\end{equation}}

\newcommand{\chiPT}{$\chi$PT }
\newcommand{\QchiPT}{Q$\chi$PT }

\newcommand{\rsquared}{\langle r^2 \rangle}

\begin{document}

\preprint{
\vbox{
\hbox{ADP-06-01/T632}
\hbox{JLAB-THY-06-01}
\hbox{Edinburgh 2006/01}
}}

\title{Strange Electric Form Factor of the Proton}

\author{D.~B.~Leinweber}
\author{S.~Boinepalli}
\author{A.~W.~Thomas$^\dagger$}
\author{P.~Wang}
\author{A.~G.~Williams}
\author{R.~D.~Young$^\dagger$}
\author{J.~M.~Zanotti$^*$}
\author{J.~B.~Zhang}
\affiliation{Special Research Centre for the
Subatomic Structure of Matter,
and Department of Physics,
University of Adelaide, Adelaide SA 5005, Australia}
\affiliation{$^\dagger$ Jefferson Laboratory, 12000 Jefferson Ave.,
Newport News, VA 23606 USA}
\affiliation{$^*$ School of Physics, University of Edinburgh,
  Edinburgh EH9 3JZ, UK} 
\begin{abstract}
By combining the constraints of charge symmetry with new chiral
extrapolation techniques and recent low-mass quenched lattice QCD
simulations of the individual quark contributions to the electric
charge radii of the baryon octet, we obtain an accurate determination
of the strange electric charge radius of the proton.  While this
analysis provides a value for $G_E^s(Q^2=0.1 {\rm GeV}^2)$ in
agreement with the best current data, the theoretical error is
comparable with that expected from future HAPPEx results from JLab.
Together with the earlier determination of $G_M^s$, this result
considerably constrains the role of hidden flavor in the structure of
the nucleon.

\end{abstract}


\maketitle

One of the great challenges of modern hadron physics is to unravel the
precise role of hidden flavors in the structure of the nucleon.
Because of their relatively light mass, strange quarks are expected to
play the biggest role and it is with respect to strangeness that there
has recently been enormous experimental progress.  Indeed, new results
on strangeness in the nucleon have been reported recently from the
HAPPEx~\cite{HAPPEx} and G0~\cite{Armstrong:2005hs} Collaborations at
JLab, which complement earlier work at MIT-Bates~\cite{Spayde:2003nr}
and Mainz~\cite{MAINZ}.  The situation is by far the best at
$Q^2=0.1\gev^2$, where in addition to the usual linear combination of
electric and magnetic form factors, a measurement of parity violation
on $^4$He allowed an accurate extraction of $G_E^s$, namely $G_E^s(Q^2
= 0.1\gev^2) = -0.013 \pm 0.028$.  A new experimental
investigation, in which this error is expected to be reduced by roughly
a factor of two, was conducted in late 2005.  This makes it imperative
to find, as we report, a way to make a theoretical estimate of
comparable accuracy --- or better.

With respect to the strange magnetic from factor of the proton we
recently reported a calculation an order of magnitude more precise
than current experiments \cite{Leinweber:2004tc,Leinweber:2005bz}.
This calculation exploited the advances in lattice QCD which have
enabled quenched QCD (QQCD) simulations of magnetic moments at pion
masses as low as 0.3--0.4
GeV~\cite{Zanotti:2001yb,FLICscaling,FLIClqm}, as well as the
development of new chiral extrapolation
techniques~\cite{FRR,Young:2004tb}. To minimize theoretical
uncertainty, an essential input was the precise (experimental)
measurements of the magnetic moments of the ground state hyperons ---
a luxury unfortunately not available for charge
radii. Nevertheless, we show here that even the limited data on
hyperon charge radii, when combined with new lattice simulations and
chiral extrapolation techniques, yield a precision commensurate with
the published data. Alternatively, using the best estimates for the
charge radii of the valence $u$ and $d$ quarks, extracted from the
lattice simulations, yields a determination of the strange quark
contribution to the proton charge radius with an uncertainty
comparable with that anticipated from the latest HAPPEx measurement.

\begin{figure}[tbp]
{\includegraphics[height=3.3cm,angle=90]{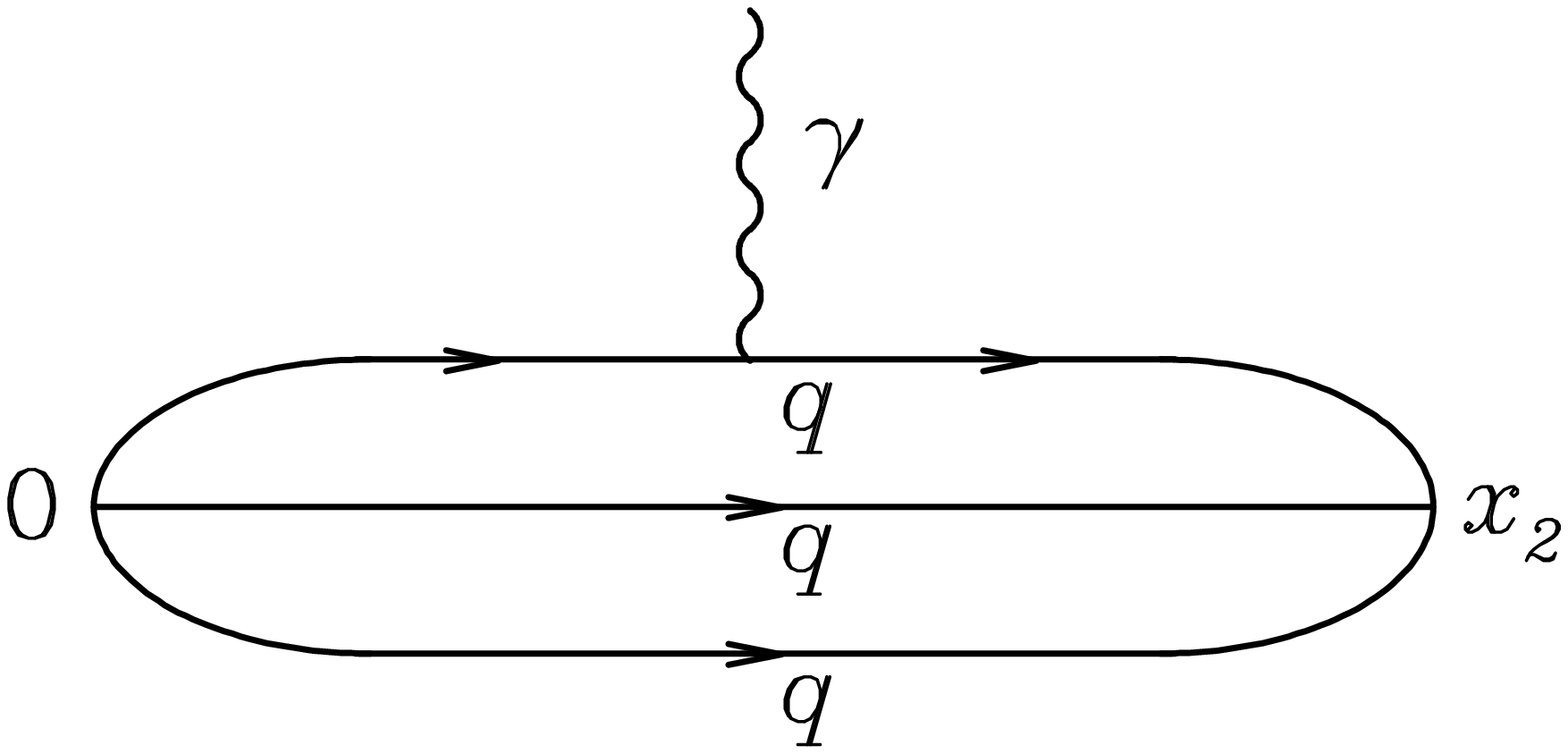} \hspace{0.8cm}
 \includegraphics[height=3.3cm,angle=90]{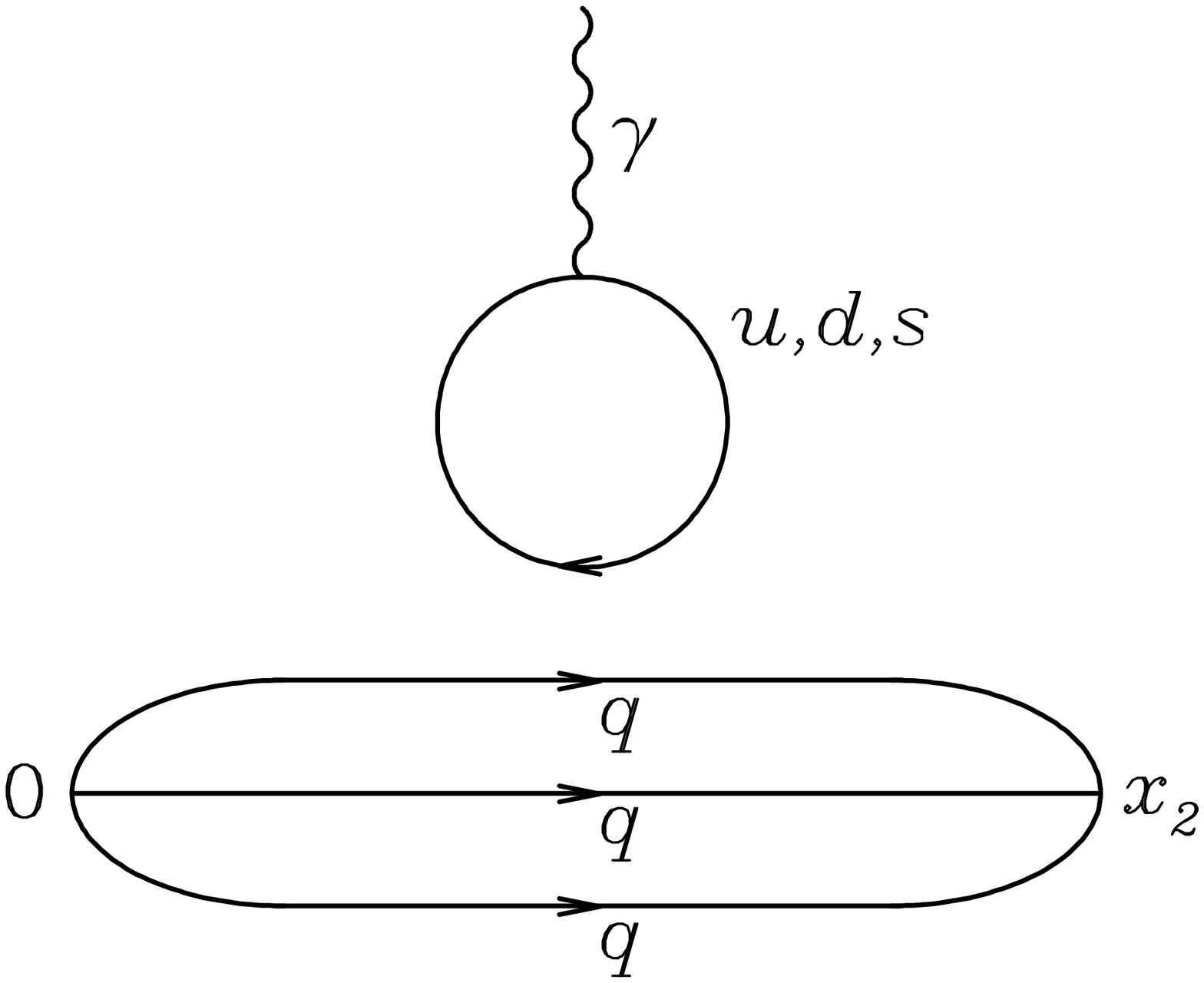}}
\caption{Diagrams illustrating the two topologically different
insertions of the current within the framework of lattice QCD.  
}
\label{topology}
\end{figure}

As illustrated in Fig.~\ref{topology}, the three point function
required to extract an electromagnetic form factor 
in lattice QCD involves two
topologically distinct processes (each incorporating an arbitrary
number of gluons and quark loops).  The left-hand diagram illustrates
the connected insertion of the current to one of the ``valence''
quarks of the baryon.  In the right-hand diagram the external field
couples to a quark loop. The latter process, where the loop involves
an $s$-quark, is entirely responsible for $G_E^s$.

Charge symmetry, which is the invariance of the QCD Hamiltonian under
rotations by 180 degrees about the 2-axis in isospace, is typically
respected at a level better than 1\% in nuclear
physics~\cite{ChargeSymm}.  This gives one confidence in using it to
extract the strange form factors experimentally and we shall also use
it.  Under charge symmetry, the electric or magnetic form factors of
the octet baryons satisfy~\cite{Leinweber:1996ie}:
\begin{equation}
\begin{array}{rcl}
p &=& e_u\, u^p + e_d\, d^p + O_N  \, , \\
n &=& e_d\, u^p + e_u\, d^p + O_N  \, , \\
\Sigma^+ &=& e_u\, u^{\Sigma} + e_s\, s^\Sigma + O_\Sigma  \, ,  \\
\Sigma^- &=& e_d\, u^{\Sigma} + e_s\, s^\Sigma + O_\Sigma  \, . 
\end{array}
\label{equalities}
\end{equation}
Here, $p$, $\Sigma^-$ etc, represent any form factor of the physical
proton and $\Sigma^-$, and similarly for the other baryons. For the
present application we take them to be the mean-square charge radii,
defined as $-6\,d/dQ^2\,G_E(Q^2)|_{Q^2=0}$.  
Similarly, $u^p$ denotes the contribution to the proton mean-square
charge radius from the two valence $u$-quarks (for $u$-quarks of unit
charge), see LHS of Fig.~\ref{topology}.

Of course, the strange quark contribution is entirely contained in 
the quark-loops, $O_N$, illustrated on the right hand side 
of Fig.~\ref{topology}. However, one has to separate the contributions 
associated with $u$  and $d$ quarks from those involving $s$ quarks.
To this end, we define $^{\ell}u$, $^{\ell}d$ and $^{\ell}s$ as the 
mean square charge radii of the loop contributions associated with 
$u$, $d$ and $s$ quarks (of unit charge), respectively. Hence we may 
write:
\begin{equation}
O_N = \frac{2}{3} \,{}^{\ell}u - \frac{1}{3} \,{}^{\ell}d -
\frac{1}{3} \,{}^{\ell}s
= \frac{{}^{\ell}s}{3} \left ( \frac{1 -
{}^{\ell}R_d^s}{{}^{\ell}R_d^s } \right ) \, , 
\label{OGEs}
\end{equation}
where the ratio of $s$- and $d$-quark {\it loops}, ${}^{\ell}R_d^s
\equiv {{}^{\ell}s}/{{}^{\ell}d}$, is expected to lie in the range
(0,1).  Note that, in deriving Eq.~(\ref{OGEs}), we have used charge
symmetry (or $m_u=m_d$) to set ${}^{\ell}u = {}^{\ell}d$
\cite{Leinweber:1996ie}.  Since the chiral coefficients for the $d$
and $s$ loops in the RHS of Fig.~1 are identical, the main difference
comes from the mass of the $K$ compared with that of the $\pi$.

With a little algebra $O_N$, and hence ${}^{\ell}s$, 
may be isolated
{}from Eqs.~(\ref{equalities}) and (\ref{OGEs}):
\be
{}^{\ell}s = \left ( {\,{}^{\ell}R_d^s \over 1 - \,{}^{\ell}R_d^s }
\right ) \left [ 2 p + n - {u^p \over u^{\Sigma}} \left ( \Sigma^+ -
\Sigma^- \right ) \right ] \, .
\label{GEsSigma} \\
\ee
It is worth emphasizing that this expression is an exact consequence
of QCD, under the assumption of charge symmetry.

There is a relation similar to Eq.~(\ref{GEsSigma}) involving the
charge radii of the $\Xi$ baryons and the ratio of the charge radii of
$u$-quarks in the $n$ and $\Xi^0$, but as there is no experimental
data at all for the cascades we do not show it.  Even in the case of
the $\Sigma$ hyperons, there is data only for the $\Sigma^-$ --- from
the SELEX Collaboration at Fermilab~\cite{SELEX}.  Contrasting the
magnetic moments, the experimental error is relatively large,
$\Sigma^- = -0.61 \pm 0.12 \pm 0.09$ fm$^2$.  Nevertheless, as we
shall see, even this is enough, in combination with lattice QCD
estimates of the ratio of charge radii in the $\Sigma^+$ compared with
the $\Sigma^-$, to obtain an accurate estimate of the strangeness
radius.

\begin{figure}[tbp]
\begin{center}
{\includegraphics[height=7.8cm,angle=90]{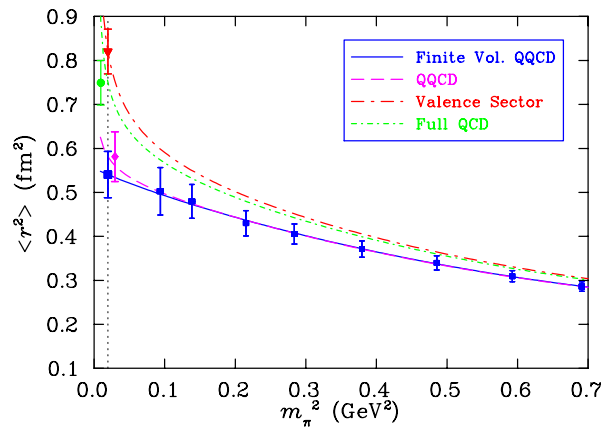}}
\vspace{-15pt}
\end{center}
\caption{Mean square charge radius of the $\Sigma^+$, as a function of
the pion mass squared, calculated in quenched lattice
QCD~\protect\cite{SOURCE}.  The solid curve displays the finite volume
fit, the infinite volume limit of quenched QCD is shown by the dashed
curve, the valence sector of full QCD is represented by the
long-dash-dot curve and the short-dash-dot curve shows the total
result, including the disconnected loops.}
\label{fig:SigmaP}
\end{figure}

\begin{figure}[tbp]
\begin{center}
{\includegraphics[height=7.8cm,angle=90]{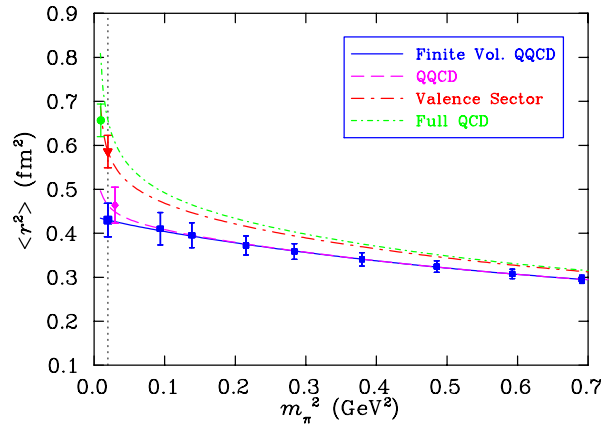}}
\vspace{-15pt}
\end{center}
\caption{Absolute value of the mean square charge radius of the
$\Sigma^-$, as a function of pion mass squared, calculated in
quenched lattice QCD~\cite{SOURCE}.  Curve descriptions are the same as
those in Fig.~\ref{fig:SigmaP}.}
\label{fig:SigmaM}
\end{figure}

\begin{figure}[tbp]
\begin{center}
{\includegraphics[height=7.8cm,angle=90]{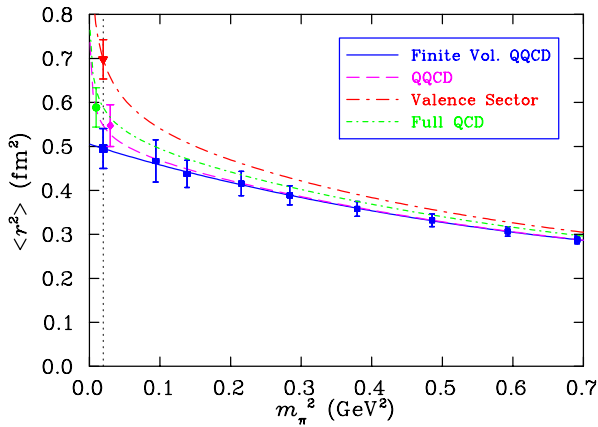}}
\vspace{-15pt}
\end{center}
\caption{Mean square charge radius of a single $u$ quark (of unit
  charge) in the proton, calculated in quenched lattice
  QCD~\cite{SOURCE}.  Curve descriptions are the same as those in
  Fig.~\ref{fig:SigmaP}.}
\label{fig:uProt}
\end{figure}

\begin{figure}[tbp]
\begin{center}
{\includegraphics[height=7.8cm,angle=90]{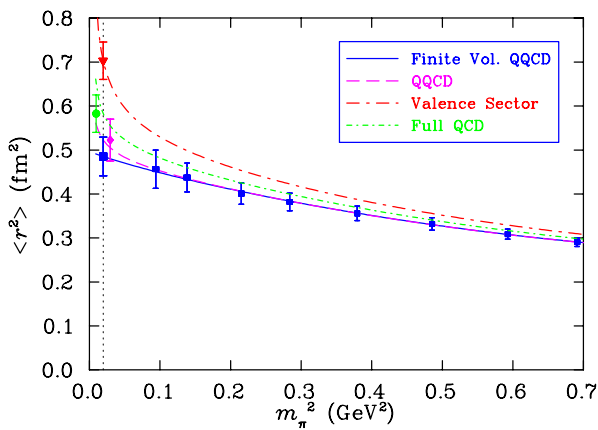}}
\vspace{-15pt}
\end{center}
\caption{Mean square charge radius of a single $u$ quark (of unit
  charge) in the $\Sigma^+$, calculated in quenched lattice
  QCD~\cite{SOURCE}.  Curve descriptions are the same as those in
  Fig.~\ref{fig:SigmaP}.}
\label{fig:uSig}
\end{figure}

Combining the experimentally measured mean square charge radii of the 
$p$ and $n$ ($0.757 \pm 0.014\fm^2$ and $-0.116 \pm 0.002\fm^2$, 
respectively~\cite{PDG}) 
we find:
\begin{eqnarray}
{}^{\ell}s &=& 
\left ( {\,{}^{\ell}R_d^s \over 1 - \,{}^{\ell}R_d^s } \right ) 
\times \nonumber \\
&& \hspace*{-5mm} \left [
1.398 - {u^p \over u^{\Sigma}} \left ( 1 + \left | \frac{\Sigma^+}{\Sigma^-}
\right |   
\right ) 0.61 \pm 0.12 \pm 0.09 \right ] .
\label{ok} 
\end{eqnarray}
To obtain $O_N$, the necessary input from lattice QCD is now the two
ratios in Eq.(\ref{ok}), namely $u^p/u^\Sigma$ and
$\Sigma^+/\Sigma^-$. Figures \ref{fig:SigmaP}, \ref{fig:SigmaM},
\ref{fig:uProt} and \ref{fig:uSig} show the state of the art
calculations from which we can extract these ratios.  The numerical
simulations of the electromagnetic form factors presented here are
carried out using the Fat Link Irrelevant Clover (FLIC) fermion action
\cite{Zanotti:2001yb} in which the irrelevant operators, introduced to
remove fermion doublers and lattice spacing artifacts, are constructed
with APE smeared links~\cite{ape}.  Perturbative renormalizations are
small for smeared links and the mean-field improved coefficients used
here are sufficient to remove ${\mathcal O}(a)$ errors from the
lattice fermion action~\cite{Bilson-Thompson:2002jk}.

The ${\mathcal O}(a)$-improved conserved vector current
\cite{Martinelli:ny} is used.  Nonperturbative improvement is achieved
via the FLIC procedure, where the terms of the Noether current having
their origin in the irrelevant operators of the fermion action are
constructed with mean-field improved APE smeared links.  The results
presented here are obtained using established
techniques~\cite{dblOctet} from a sample of 400, $20^3 \times 40$
mean-field ${\cal O}(a^2)$-improved Luscher-Weisz
\cite{Luscher:1984xn} gauge field configurations, having a lattice
spacing of 0.128 fm, determined by the Sommer scale $r_0=0.50$ fm.
Charge radii are obtained from dipole fits to quark-sector electric
form factors, calculated at $Q^2 = 0.227 \pm 0.002\gev^2$. The dipole
form is well supported physically and in lattice QCD simulations over
a range of quark masses \cite{Gockeler:2003ay}.

From Figs.~\ref{fig:SigmaP} and \ref{fig:SigmaM} we see that the
radius of the $\Sigma^+$ grows significantly larger than that of the
$\Sigma^-$ as we approach the chiral limit. This has a relatively
simple, physical interpretation in terms of the more extended spatial
distribution of light quarks compared with strange. These simulation
results are fit using finite-range regularised (FRR), quenched chiral
effective field theory formulated on a finite volume. Upon extracting the
fit, one can restore the infinite-volume limit within the effective
field theory --- where corrections are observed to be quite small
right down to the lightest simulated pion mass. The QCD valence sector
and total QCD (including loop insertions) contributions are then
estimated by replacing the long-range ``tail'' of \QchiPT by the
corresponding \chiPT ``tail'' in the same fashion as
Refs.~\cite{Leinweber:2004tc,Young:2004tb} --- which extends upon the
empirical success observed in Ref.~\cite{Young:2002cj}.

Using the techniques described in Ref.~\cite{Leinweber:2004tc}, we
estimate the effects of the dynamical sea and extrapolate the results
to the physical quark mass.  The ratio of the $\Sigma$ charge radii is
found to be $|\Sigma^+/\Sigma^-| = 1.141 \pm 0.036 \pm 0.010$ ---
where we observe the statistical fluctuations (first error) have been
significantly reduced by constructing the ratio. The second
uncertainty arises from the FRR scale dependence,
$\Lambda=0.8\pm0.1\gev$, constrained to provide agreement with
experiment where available.

In Figs.~\ref{fig:uProt} and \ref{fig:uSig} we show the corresponding
analysis for the $u$-quark mean square radii in the proton and
$\Sigma^+$, respectively.  Here we observe that there is very little
sensitivity to the mass of the spectator quark, with the ratio of the
extrapolated valence contributions giving our best estimate of
$u^p/u^\Sigma = 0.993 \pm 0.048 \pm 0.000$.  The latter uncertainty
indicates negligible sensitivity to the FRR scale.  Putting these best
estimates into Eq.~(\ref{ok}), we obtain ${}^{\ell}s =
\Rlsd(1-\Rlsd)^{-1}[0.11 \pm 0.06 \pm 0.02 \pm 0.33 \fm^2]$,
where the uncertainties are respectively, statistical, systematic and
experimental.

In order to complete the evaluation, we also need to estimate the
ratio of disconnected $s$ to light quark loops, ${}^{\ell}R_d^s$. No
attempt has yet been made to evaluate this ratio in an {\it ab initio}
lattice QCD simulation, although future work could attempt this by
building on the work of Refs.~\cite{kentucky,Lewis:2002ix}.  We follow
the same procedure as in Ref.~\cite{Leinweber:2004tc}, but now for the
charge radii rather than the magnetic moments, using the relative
strength of the FRR loop contributions.  With the FRR scale varying
over the generous range $0.8 \pm 0.2\gev$, this provides
${}^{\ell}R_d^s = 0.16 \pm 0.04$.

The final result for the proton strange, mean square charge radius is
then $\protect{\rsquared^p_s}= - {}^{\ell}s/3 = -0.007 \pm 0.004 \pm
0.002 \pm 0.021\fm^2$, 
where the uncertainties are statistical, FRR scale and experimental in
origin.  
Lattice scale determination uncertainties are negligible.
At $Q^2=0.1\gev^2$, we expect the first term in the expansion
of the electric form factor to dominate, $G_E(Q^2)=\protect{-Q^2
\rsquared/6}$. The strangeness electric form factor, conventionally
defined without the charge factor, is then found to be
\begin{equation}
G_E^s(0.1 \, {\rm GeV}^2) = -0.009 \pm 0.005 \pm 0.003 \pm 0.027\,.
\end{equation}
This is consistent with the analysis of world data at $0.1\gev^2$,
namely $G_E^s(0.1\gev^2) = -0.013 \pm 0.028$. The error on this
theory calculation is of similar precision to the experiment.

It is interesting that a combination of data from the SELEX
collaboration at Fermilab, together with the constraints of charge
symmetry and modern lattice QCD can provide a tight constraint on the
strange electric form factor of the proton.  For the present time the
errors are dominated by the experimental errors in the knowledge of
the $\Sigma^-$ charge radius.  It would clearly be valuable to have
new and more accurate measurements on both the $\Sigma^-$ and
$\Sigma^+$ hyperons. Data on the $\Xi^-$ (and $\Xi^0$) would, as in
the case of the strangeness magnetic moment, provide a valuable
additional constraint.

However, noting that the uncertainty in our result is dominated by the
lack of precision in the observed $\Sigma^-$ charge radius, we also
consider a somewhat more ambitious approach. That is, we remove the
dependence on the hyperons altogether and require only the observed
nucleon charge radii and our best estimate for the quark-sector
connected-current insertion, either $u^p$ or $d^p$,
\begin{eqnarray}
{}^{\ell}s &=& \left ( \frac{{}^{\ell}R_d^s}{1 - \,{}^{\ell}R_d^s} \right)
             \left[ 2 p + n - u^p \right] \, ,\\
           &=& \left ( \frac{{}^{\ell}R_d^s}{1 - \,{}^{\ell}R_d^s} \right)
             \left[ p + 2 n - d^p \right] \, .
\end{eqnarray}
In this case, uncertainties in the lattice scale determination,
$a=0.128(6)$ fm,
\cite{Leinweber:2005bz} become significant.
Our best extrapolation of the valence sector contributions to the mean
square charge radii in full QCD gives
$u^p=1.396 \pm 0.090 \pm 0.046 \pm 0.086\ \fm^2$ and
$d^p=0.553 \pm 0.063 \pm 0.010 \pm 0.047\ \fm^2$, 
where the uncertainties are statistical, FRR scale and lattice scale
in origin.  These values provide two independent evaluations of the
strangeness mean-square charge radius,
$\protect{\rsquared^p_s}= 0.000 \pm 0.006 \pm 0.007\ {\rm fm}^2$ and
$\protect{\rsquared^p_s}= 0.002 \pm 0.004 \pm 0.004\ {\rm fm}^2$, 
where the first uncertainty is the combined statistical and
experimental errors, and the second is the combined FRR and lattice
scale uncertainty.
Combining these two in the extraction of the form factor at
$Q^2=0.1\gev^2$ gives
\begin{equation}
G_E^s(0.1 \, {\rm GeV}^2) = +0.001 \pm 0.004 \pm 0.004 \,.
\end{equation}
\begin{table}
\caption{Results (fm${}^2$) are compared with experiment.
  Uncertainties are statistical and systematic respectively. Note that
  $\rsquared_s^p$ is reported including the strange-quark charge.}
\label{tab:comp}
\begin{ruledtabular}
\begin{tabular}{lll}
Quantity          & This Work                      & Experiment  \\
\hline
$\rsquared^p$     & $+0.685 \pm 0.047 \pm 0.051$   &$+0.757\pm 0.014$  \\
$\rsquared^n$     & $-0.158 \pm 0.029 \pm 0.016$   &$-0.116\pm 0.002$  \\
$\rsquared^{\Sigma^-}$   
                  & $-0.657 \pm 0.037 \pm 0.045$   &$-0.610\pm 0.150$  \\
$\rsquared_s^p$   & $+0.001 \pm 0.004 \pm 0.004$   &$-0.010\pm 0.022$
\end{tabular}
\end{ruledtabular}
\vspace{-10pt}
\end{table}
Following the procedure outlined in Ref.~\cite{Thomas:2005qb},
charge-symmetry violations (CSV) of the maximum size estimated
theoretically \cite{Miller:1997ya} are found to be small with respect
to the error on the experimental proton mean-square radius.  The small
magnitude of the strange form factors, predicted here and in
Ref.~\cite{Leinweber:2004tc}, suggests the experimental signal may
be similar in size to the CSV effects. As the precision in
experimental programs are improved a more careful QCD-analysis of CSV
effects will be required.

Table~\ref{tab:comp} provides a comparison of the present calculation
and experiment.
The quoted errors on the present prediction are comparable to, {\it or
even better than}, those anticipated in the forthcoming HAPPEx
results.

We thank the Australian Partnership for Advanced Computing (APAC) and
the South Australian Partnership for Advanced Computing (SAPAC) for
supercomputer support enabling this project.  This work is
supported by the Australian Research Council and by DOE contract
DE-AC05-84ER40150, under which SURA operates Jefferson Laboratory.

\end{document}